\documentclass[aps,prl,twocolumn,showpacs]{revtex4-1}
\usepackage{amssymb}
\usepackage{graphicx}

\begin{document}

\title{Soliton gyroscopes in media with spatially growing repulsive
nonlinearity}
\date{\today}
\author{Rodislav Driben$^{1,2}$}
\author{Yaroslav V. Kartashov$^{3,4}$}
\author{Boris A. Malomed$^2$}
\author{Torsten Meier$^1$}
\author{Lluis Torner$^3$}
\affiliation{$^1$Department of Physics \& CeOPP, University of
Paderborn, Warburger Str. 100,
Paderborn D-33098, Germany \\
$^2$Department of Physical Electronics, School of Electrical
Engineering, Faculty of Engineering,
Tel Aviv University, Tel Aviv 69978, Israel\\
$^3$ICFO-Institut de Ciencias Fotoniques, and Universitat
Politecnica de Catalunya,
Mediterranean Technology Park, E-08860 Castelldefels (Barcelona), Spain\\
$^4$Institute of Spectroscopy, Russian Academy of Sciences, Troitsk,
Moscow, 142190, Russia}

\begin{abstract}
We find that the recently introduced model of self-trapping
supported by a spatially growing strength of a repulsive
nonlinearity gives rise to robust vortex-soliton tori, i.e.,
three-dimensional vortex solitons, with topological charges $S\geq
1$. The family with $S=1$ is completely stable, while the one with
$S=2$ has alternating regions of stability and instability. The
families are nearly exactly reproduced in an analytical form by the
Thomas-Fermi approximation (TFA). Unstable states with $S=2$ and $3$
split into persistently rotating pairs or triangles of unitary
vortices. Application of a moderate torque to the vortex torus
initiates a persistent precession mode, with the torus' axle moving
along a conical surface. A strong torque heavily deforms the vortex
solitons, but, nonetheless, they restore themselves with the axle
oriented according to the
vectorial addition of angular momenta. %Thus, these vortex tori are
%exceptionally robust objects. The model is relevant to Bose-Einstein
%condensates (BECs), and to field theories dealing with complex
%multidimensional topological modes.
\end{abstract}

\maketitle

The creation of multidimensional spatiotemporal solitons is a challenging
problem of nonlinear physics, including optics and matter-wave dynamics \cite%
{review1}-\cite{review2}. Fundamental multidimensional solitons are prone to
instabilities caused by the beam collapse \cite{Kuznetsov}, while vortex
solitons \cite{Anton} are highly vulnerable to azimuthal modulational
instabilities that split them into fragments \cite{PeliKivetal}. A related
topic is search for various complex topological states, such as knots,
\textquotedblleft Q-balls", \textquotedblleft vortons", and skyrmions, in
classical-field systems \cite{fields,fields-review}. Those states model
topological excitations in ferromagnets \cite{ferro-vortring},
superconductors \cite{superconductor-knot}, Bose-Einstein condensates (BECs)
\cite{BEC-skyrm}, and in barionic matter in the low-energy limit \cite%
{low-energy}.

A few methods for the stabilization of multidimensional solitons and
vortices have been put forward theoretically. These include the use
of competing (cubic-quintic \cite{Manolo}-\cite{VA-CQ} or
quadratic-cubic \cite{QC}) nonlinearities, trapping configurations \cite%
{trap}, and, as demonstrated experimentally \cite{lattice0} and
theoretically \cite{lattice}, periodic lattice potentials. The stabilization
may be enhanced by \textquotedblleft management" techniques, i.e., periodic
alternation of the sign of the nonlinear term \cite{management}. Another
possibility for the stabilization is provided by the use of nonlocal
nonlinearity \cite{nonlocal}.

A new approach to the creation of self-trapped fundamental and vortex
solitons was recently introduced in Refs. \cite{we1,we2} and elaborated in
diverse settings \cite{further}: a repulsive (defocusing) nonlinearity,
whose local strength in space of dimension $D$ grows from the center to the
periphery at any rate faster than $r^{D}$ ($r$ is the radial coordinate),
supports remarkably robust families of fundamental and vortical solitons for
$D=1$ and $2$. Such type of the nonlinearity modulation may be generated by
means of several techniques. In optical media, one may use inhomogeneous
doping \cite{Kip}. %Uniformly distributed
%dopants can be used too, if an external field induces a spatial
%modulation of the detuning in them, decreasing from the center to
%periphery.
In BECs, the tunability of the magnetic Feshbach resonance (FR) \cite%
{extreme-tunability} allows the creation of spatially inhomogeneous
nonlinearity landscapes by means of properly shaped magnetic fields \cite%
{Feshbach,review1}. Furthermore, optically controlled FR \cite%
{optical-Feshbach}, as well as combined magneto-optical control mechanisms
\cite{magnetic-optical}, make it possible to create a diverse set of spatial
profiles of the self-repulsive nonlinearity. In particular, the required
pattern of the laser-field intensity controlling the optically induced FR
can be \textit{painted} in space, as demonstrated in Ref. \cite{paint}. In
terms of the above-mentioned classical field-theory models, a spatially
growing component may play the role of an inhomogeneous nonlinearity
coefficient for another component, coupled to the former one by a repulsive
quintic interaction.

To date, self-trapping by means of the repulsive-nonlinearity scheme
was studied only for to one- and two-dimensional (1D and 2D)
geometries, except for the simplest case of 3D spherically symmetric
solitons, which were mentioned in Ref. \cite{we2}. The objective of
the present Letter is to construct stable 3D vortex tori with
various topological charges $S$. This is a challenging goal. Thus
far, limited results for stable 3D vortex tori with $S=1$ were
obtained solely in models with uniform cubic-quintic \cite{CQ} and
quadratic-cubic \cite{QC} focusing-defocusing nonlinearities. Only
very broad vortex solitons may be stable against splitting in these
systems, which makes the possibility of their experimental
realization unlikely, and so far no stable 3D vortices with $S>1$
have been found. Here we show that the family of vortex-solitons
tori with $S=1$ is completely stable, while the families with $S=2$
and $3$ feature instability segments, with the unstable tori
splitting into pairs or triangles of mutually orbiting unitary
vortices. In spite of the apparent complexity of the 3D model and
vortex solitons sought for, families of solutions with $S=0$ and
$S=1,2$ are obtained below in very accurate explicit forms by means
of the variational \cite{AblLis,VA-BEC} and Thomas-Fermi
\cite{TF-Pu} approximations (VA and TFA), respectively.

The 3D vortex-soliton tori may be viewed as matter-wave/field-theory
counterparts of mechanical gyroscopes, whose most remarkable dynamical
property is that application of a torque drives them into precession \cite%
{LL}. We find that the vortex tori in the present model behave as precessing
gyroscopes under the action of a moderate torque. To the best of our
knowledge, no similar effect has ever been shown for 3D vortex solitons.
Furthermore, due to the robustness of the vortex-soliton tori in the setting
considered here, we find that they may restore their shapes even after
significant deformations caused by the application of a strong torque.

We describe the evolution of 3D matter-wave excitations by the
Gross-Pitaevskii equation for the scaled wave function $q$:
\begin{equation}
i\frac{\partial q}{\partial t}=-\frac{1}{2}\nabla ^{2}q+\sigma (r)\left\vert
q\right\vert ^{2}q,  \label{1}
\end{equation}%
where Laplacian $\nabla ^{2}$ acts on coordinates $\left( x,y,z\right) $%
, with $r^{2}=x^{2}+y^{2}+z^{2}$, and the anti-Gaussian isotropic modulation
of the local self-repulsion is chosen, $\sigma (r)=\exp \left(
r^{2}/2\right) $, which helps us to present the results in a compact form,
although, as said above, the necessary condition for self-trapping of
finite-norm modes in the 3D case is that $\sigma (r)$ must grow at $%
r\rightarrow \infty $ at any rate faster than $r^{3}$. Equation
(\ref{1}) conserves the Hamiltonian and norm, $H=(1/2\int \int \int
\left[ \left\vert \nabla q\right\vert ^{2}+\sigma (r)|q|^{4}\right]
dxdydz$ and $N=\int \int \int \left\vert q\left( x,y,z,t\right)
\right\vert ^{2}dxdydz$, along with the vector of the angular
momentum ($q^{\ast }$ is for the complex conjugate):
\begin{equation}
\mathbf{M}=-i\int \int \int q^{\ast }\left( \mathbf{r}\times \nabla \right)
q\left( x,y,z\right) dxdydz.  \label{M}
\end{equation}

In cylindrical coordinates $\left( \rho ,z,\theta \right) $, stationary
solutions with chemical potential $\mu $ and integer vorticity $S\geq 0$ are
looked for as $q=\exp \left( -i\mu t+iS\theta \right) u\left( \rho ,z\right)
$, where the real function $u\left( \rho ,z\right) $ satisfies
\begin{equation}
\mu u=-\frac{1}{2}\left( \frac{\partial ^{2}u}{\partial \rho ^{2}}+\frac{1}{%
\rho }\frac{\partial u}{\partial \rho }+\frac{\partial ^{2}u}{\partial z^{2}}%
-\frac{S^{2}u}{\rho ^{2}}\right) +e^{\left( \rho ^{2}+z^{2}\right) /2}u^{3}.
\label{phi}
\end{equation}%
The Lagrangian corresponding to Eq. (\ref{phi}) is $L=2\pi \int_{-\infty
}^{+\infty }dz\int_{0}^{\infty }\rho d\rho \mathcal{L}$, with density $%
\mathcal{L}=%\left( 1/4\right) \left[
-2\mu u^{2}+\left( \partial u/\partial \rho \right) ^{2}+\left( \partial
u/\partial z\right) ^{2}+S^{2}\rho ^{-2}u^{2}+e^{\left( \rho
^{2}+z^{2}\right) /2}u^{4}%\right]
$. Families of stationary solutions are characterized by the
dependence of the norm, $N$, on $\mu $, while the angular momentum
and norm of the solutions are related by
\begin{equation}
M_{z}=SN,M_{x,y}=0.  \label{Mz}
\end{equation}%

The VA for solutions to Eq. (\ref{phi}) is based on the ansatz whose
functional form is suggested by matching to the modulation profile
in Eq. (\ref{phi}),
\begin{equation}
u_{\mathrm{VA}}\left( \rho ,z\right) =A\rho ^{S}\exp \left( -\left(
\rho ^{2}+z^{2}\right) /4\right).  \label{ans}
\end{equation}%
Its norm is $N_{S}^{\mathrm{VA}}=\left( 2\pi \right) ^{3/2}S!A^{2}$.
The substitution of the ansatz into the Lagrangian leads to the
variational equation, $\partial L/\partial \left( A^{2}\right) =0$,
to yield $A^{2}=\left[ S!/\left( 2S\right) !\right] \left[ \mu
-\left( 3+2S\right) /8\right] $. The comparison of ansatz
(\ref{ans}) with numerically found solutions [shown below in Fig.
\ref{fig2}(d)]
indicates that the VA predicts accurate results only for $S=0$, \textit{viz}%
.,%
\begin{equation}
N_{S=0}^{\mathrm{VA}}=\left( 2\pi \right) ^{3/2}\left( \mu -3/8\right) .
\label{0}
\end{equation}

For $S\geq 1$, accurate results are produced by the TFA, which was not
previously applied to the description of multidimensional solitons. This
approximation neglects derivatives in Eq. (\ref{phi}) (which is readily
justified for the case of strong repulsive nonlinearity \cite{TF-Pu}),
yielding%
\begin{equation}
u_{\mathrm{TFA}}^{2}=\left\{
\begin{array}{c}
0,~\mathrm{at}~~\rho ^{2}<\rho _{S}^{2}\equiv S^{2}/\left( 2\mu \right) , \\
e^{-\left( \rho ^{2}+z^{2}\right) /2}\left[ \mu -S^{2}/\left( 2\rho
^{2}\right) \right] ,~\mathrm{at}~~\rho ^{2}>\rho _{S}^{2}.%
\end{array}%
\right.   \label{uTFA}
\end{equation}%
In particular, the first line in this expression corresponds to the hole at
the center of a vortex state, see Fig. \ref{fig2}(d). The $N(\mu )$
dependence produced by Eq. (\ref{uTFA}) is%
\begin{equation}
N_{S}^{\mathrm{TFA}}=4\left( 2\pi \right) ^{3/2}\mu e^{-S^{2}/\left( 4\mu
\right) }\int_{0}^{\infty }dR\frac{e^{-R}}{4R+\left( S^{2}/\mu \right) },
\label{NTFA}
\end{equation}%
which simplifies for $S=0$:
\begin{equation}
N_{S=0}^{\mathrm{TFA}}=\left( 2\pi \right) ^{3/2}\mu ,  \label{TFA-0}
\end{equation}%
cf. Eq. (\ref{0}), and for $\mu \gg S^{2}$:
$N_{S}^{\mathrm{TFA}}(\mu )\approx \left( 2\pi \right) ^{3/2}\left[
\mu -\left( S^{2}/4\right) \ln \left( 4e\mu /S^{2}\right) \right] $.
In the general case, the integral in Eq. (\ref{NTFA}) can be
calculated numerically.

\begin{figure}[t]
\centering \centering
\includegraphics[width=8cm]{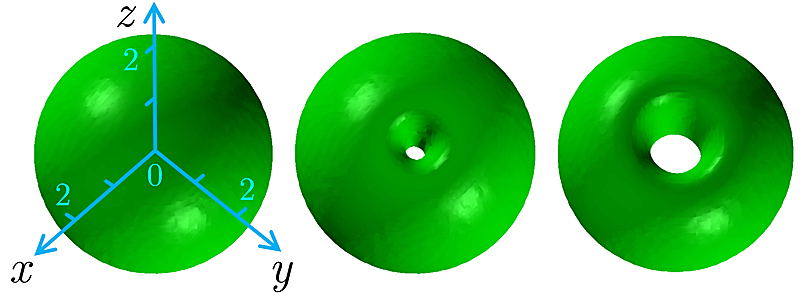}
\caption{(Color online) Isosurface plots for level $\left[ u\left(
x,y,z\right) \right] ^{2}=3$ show density distributions in stable solitons
with topological charges $S=0$, $1$, and $2$ (left, center, and right
panels, respectively), for $\protect\mu =16$. The respective norms are $%
N_{0}=246.1,~N_{1}=225.2,$ $N_{2}=191.7$.}
\label{fig1}
\end{figure}

\begin{figure}[t]
\centering \centering
\includegraphics[width=8cm]{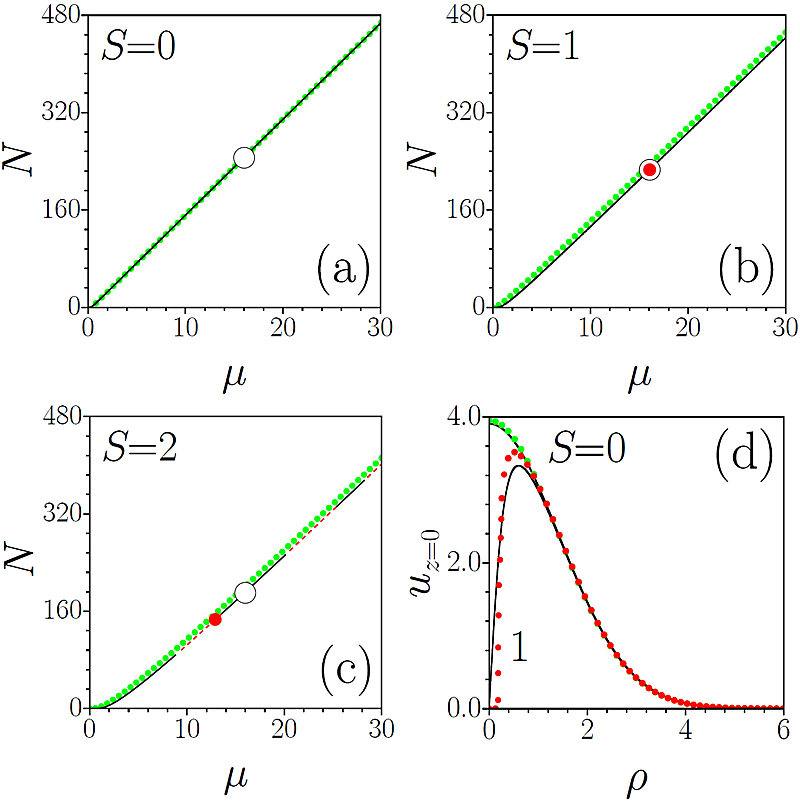}
\caption{(Color online) (a,b,c): Norm $N$ versus chemical potential $\protect%
\mu $ for the solitons with topological charges $S=0,1,2$. Solid
black and dashed red lines designate stable and unstable (for $S=2$)
branches, respectively. Green dotted lines represent the prediction
of the VA in (a),
and of the TFA in (b,c), see Eqs. (\protect\ref{0}) and (\protect\ref{NTFA}%
), respectively. White circles designate examples of stable solutions
displayed in Fig. 1. Bold red dots correspond to the vortex-soliton tori
displayed in the first two rows of Fig. \protect\ref{fig3}.(d) Comparison of
numerically computed soliton profiles $u\left( \protect\rho ,z=0\right) $
(solid lines) with profiles predicted by the VA (green dots for $S=0$) and
TFA (red dots for $S=1$) at $\protect\mu =16$.}
\label{fig2}
\end{figure}

Generic examples of numerically found stable self-trapped solutions to Eq. (%
\ref{phi}), with $S=0,1,2$ and $\mu =16$, are shown in Fig.
\ref{fig1}; families of such vortex soliton tori are represented by
$N(\mu )$ curves in Fig. \ref{fig2}(a-c); and the comparison of the
VA- and TFA-predicted solution profiles with their numerically found
counterparts is displayed in Fig. \ref{fig2}(d). The stability of
the solitons was checked by means of systematic simulations of the
evolution of perturbed solutions. The stationary and evolutionary
solutions were obtained by means of the Newton's and split-step
methods, respectively, in a 3D domain of size $20^{3}$, covered by a
mesh of $256^{3}$ points.

As seen in Fig. \ref{fig2}, the VA and TFA provide very accurate predictions
for the families with $S=0$ and $S\geq 1$, respectively (for $S=0$, the VA
is slightly more accurate than the TFA). The branches with $S=0$ and $1$
were found to be completely stable, while the one with $S=2$ (as well as its
counterpart with $S=3$, which is not shown here) features alternating
regions of stability and instability. All branches satisfy the \textit{%
anti-Vakhitov-Kolokolov} criterion, $dN/d\mu >0$, which is a necessary (but,
generally, not sufficient) condition for the stability of solitons supported
by repulsive nonlinearities \cite{anti}.

\begin{figure}[t]
\centering \centering
\includegraphics[width=8cm]{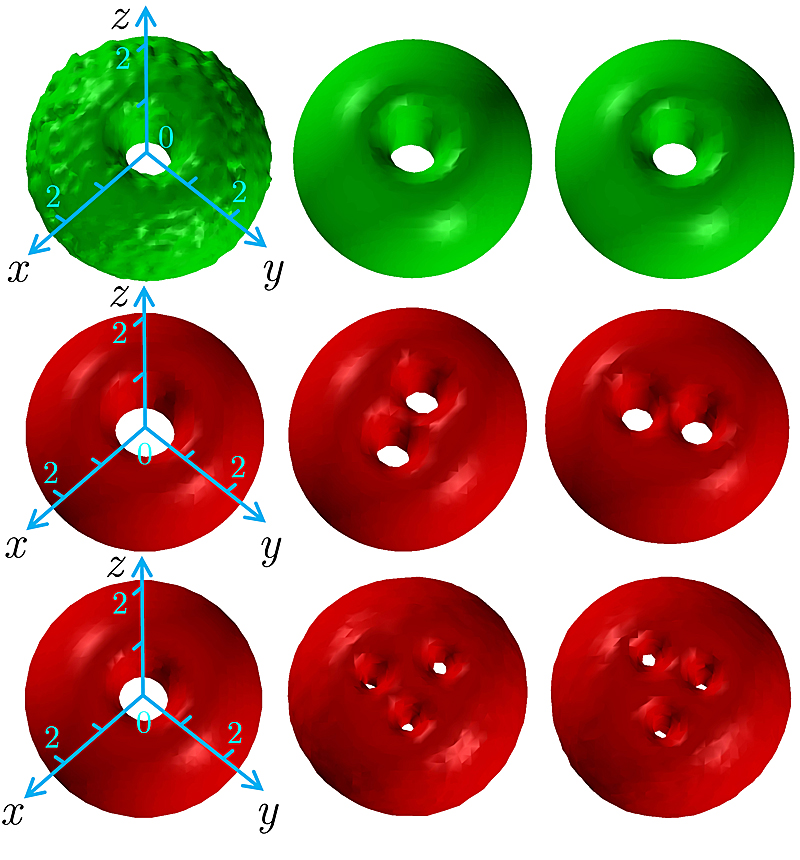}
\caption{(Color online) Top row: Stable evolution of a perturbed
vortex-soliton torus with $S=1$ and $\protect\mu =16$. Density distributions
are shown at $t=0,~450$, and $900$ (from left to right). Middle row:
Splitting of the unstable vortex soliton with $S=2$, $\protect\mu =13$ into
a steadily rotating pair of unitary vortices. Density distributions are
shown at $t=0$, $230$, and $310$. Bottom row: Splitting of the unstable
vortex-soliton torus with $S=3$, $\protect\mu =25$ into a rotating triangle
formed by unitary vortices. Density distributions are shown at $t=0,~165$
and $180$. In the top and middle rows, and in the bottom one isosurfaces
correspond to $|q\left( x,y,z\right) |^{2}=2.5$ and $|q\left( x,y,z\right)
|^{2}=3$, respectively.}
\label{fig3}
\end{figure}

Typical examples of stable and unstable evolution of perturbed
vortex-soliton tori with $S=1,2,3$ are displayed in Fig. \ref{fig3}. In the
case of unstable evolution, the phase dislocation located at the center of
the vortex tori splits into a pair (for $S=2$) or a triangle (for $S=3$) of
charge-$1$ dislocations, which rotate around the $z$-axis. The total moment (%
\ref{M}) is conserved in the course of the evolution. This
instability scenario is in contrast to those for vortex solitons in
other models with local nonlinearities, where vortex solitons split
into sets of separating vorticityless fragments \cite{PeliKivetal,
CQ, QC, Assanto-book}. The splitting of 2D higher-order solitary
vortices into separating unitary ones in a nonlocal liquid-crystal
model was demonstrated in Ref. \cite{Assanto-book} (while here the
split unitary vortices remain bound).

\begin{figure}[t]
\centering \centering
\includegraphics[width=8cm]{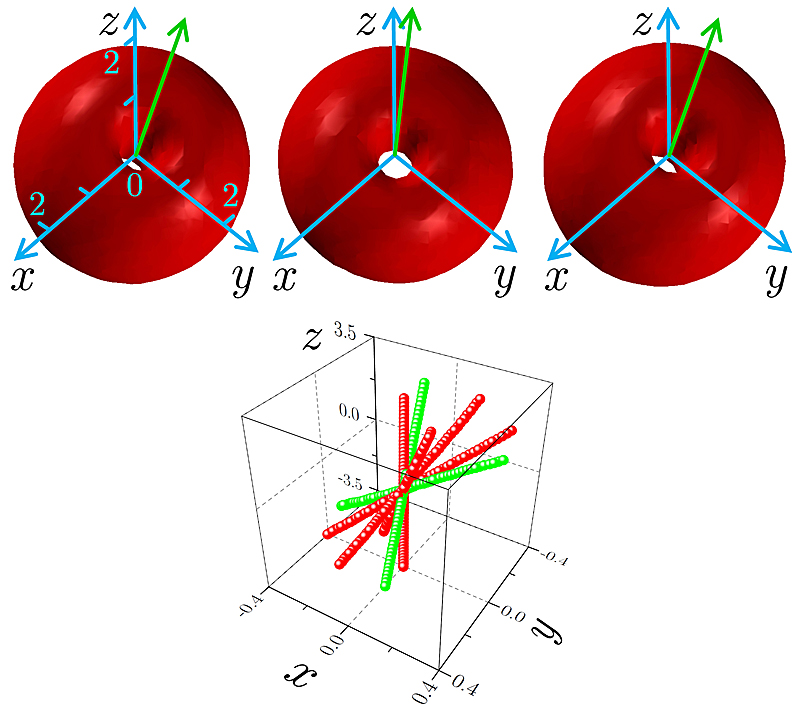}
\caption{(Color online) Precession of a vortex soliton (with $S=1$)
initiated by a relatively weak torque (\protect\ref{T}), with $\protect%
\alpha =4$ and $x_{0}=5$. Top row: Isosurface plots at $t=90,$ $90.8$, and $%
91.5$ (from left to right), time $\Delta t=1.5$ between the first and last
panels being the precession period. Green arrows indicate respective
orientations of the axle of the torus. Bottom row: Positions of the axle at
different moments of time, within one precession period. The positions shown
in green correspond to the configurations displayed in the top row.}
\label{fig4}
\end{figure}

The above results suggest that 3D vortex-soliton tori may feature
dynamics similar to that of mechanical gyroscopes. A salient
property of gyroscopes is that the application of a torque
perpendicular to the original angular momentum causes their
precessing motion \cite{LL}. The vortex solitons considered here
share this property. To elucidate it, a torque, with the angular
momentum directed along axis $y$, was applied by multiplying the
wave function of a stable vortex soliton, whose axle is directed
along $z$, by
\begin{equation}
T=\exp \left( i\alpha z\tanh \left( x/x_{0}\right) \right) .  \label{T}
\end{equation}%
When the torque is relatively small, it induces a periodic precession of the
axle of the vortex-soliton torus along a slightly deformed conical surface,
as shown in Fig. \ref{fig4}. The central axis of the cone coincides with the
direction of the total angular momentum, which is the vectorial sum of the
original momentum (\ref{Mz}) and the torque momentum corresponding to factor
(\ref{T}). The latter component can be found in an analytical form for the
case when the shape (\ref{T}) is smooth, with large $x_{0}$, and the form of
the vortex-soliton torus is taken as per Eq. (\ref{uTFA}): $M_{y}=\left(
\alpha /2x_{0}\right) \left[ N_{S}-\left( 2\pi \right) ^{3/2}\left( \mu
-S^{2}/4\right) \right] $. Note that this expression vanishes for
fundamental solitons, with $S=0$, as follows from Eq. (\ref{TFA-0}).
%In the opposite limit of the sharply shaped torque impulse, which
%corresponds to $x_{0}\rightarrow 0$, i.e., replacement of $\tanh \left(
%x/x_{0}\right) $ by $\mathrm{sgn}(x)$ in Eq. (\ref{T}), the angular momentum
%imparted by the impulse to the fundamental soliton can be found in the
%analytical form too, using either the VA or TFA in the form of Eqs. (\ref%
%{ans})\ or (\ref{uTFA}), respectively:
%\begin{equation}
%M_{y}^{(S=0)}=\alpha N_{S=0}\left( 1-1/\sqrt{\pi }\right)   \label{MyS=0}
%\end{equation}%
%(this result is identical for the VA and TFA).

%Then, relations (\ref{MyS=0}) and (\ref{Mz}) suggest that the application of
%the torque impulse to the fundamental soliton may transform it into a vortex
%torus with spin $S$ and the axle oriented parallel to $y$, provided that the
%impulse's strength exceeds the following minimum value (and assuming that
%the mode keeps the original norm):%
%\begin{equation}
%\alpha >\alpha _{S}\equiv S/\left( 1-1/\sqrt{\pi }\right) \approx 2.3S.
%\label{>}
%\end{equation}

\begin{figure}[t]
\centering \centering
\includegraphics[width=8cm]{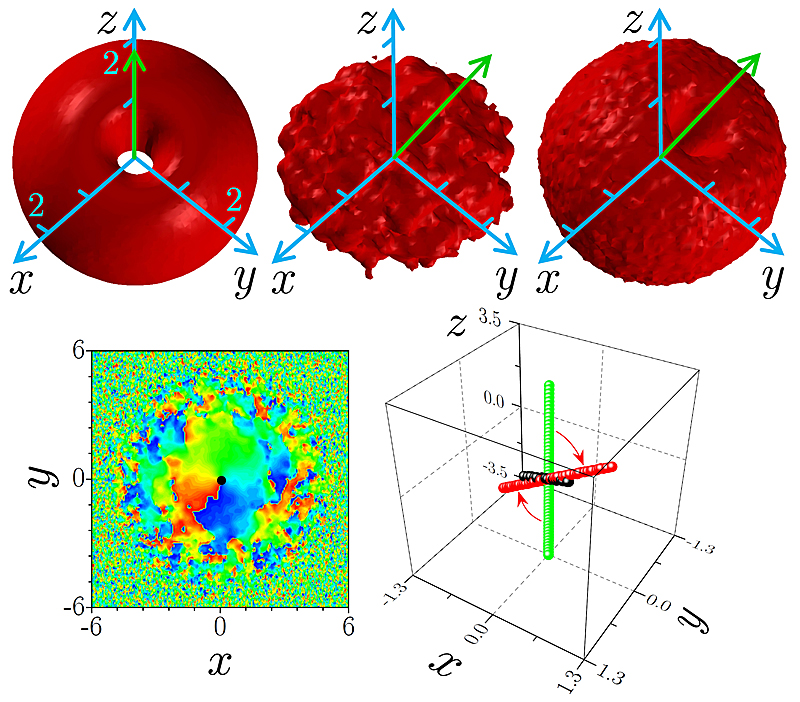}
\caption{(Color online) Top row: Isosurface plots, drawn at $\left\vert
q\left( x,y,z\right) \right\vert ^{2}=2.5$, show the violent deformation of
the vortex-soliton torus, with $S=1$ and $\protect\mu =10$, by a torque (%
\protect\ref{T}) with $\protect\alpha =10$ and $x_{0}=3$, followed by its
gradual recovery. The three plots pertain to $t=0,3,$ and $60$, from left to
right, with the green arrows indicating the respective orientations of the
axle of the torus (at $t=3$, this is actually the local orientation at the
torus center, the axle being strongly deformed in this configuration).
Bottom row, left: The phase distribution in the cross section $z=0$ at $t=3$
(left), with the bold dot showing the position of the axle. Right: arrows
show the effective rotation of the axle from its initial direction (green)
at $t=0$ to the final one at $t=60$ (red). The black segment shows a short
central fragment of the curved axle of the strongly deformed torus at $t=3$.}
\label{fig5}
\end{figure}

The precession of 3D vortex solitons, which, to the best of our knowledge,
was never shown in previous works, is analogous to the precession motion of
mechanical gyroscopes. An notable difference in the dynamics of the
vortex-soliton gyroscopes from their mechanical counterparts is the reaction
to a strong torque. A typical example is shown in Fig. \ref{fig5}, where
application of a strong torque apparently deforms the vortex torus beyond
recognition. Nevertheless, a close inspection of its phase profile at $z=0$,
displayed in the left bottom panel, and a related density profile (not shown
here) indicates that the deformed object keeps the inner vorticity, along
with the corresponding inner hole, whose environs are strongly deformed by
the perturbation. The simulations reveal that, in the course of the
subsequent evolution, the shape of the vortex-soliton torus is gradually
restored, with the straight axle aligned parallel to the total angular
momentum, which is the sum of the original momentum and that added by the
impulse. The resulting rotation of the axle from the initial to the final
position is highlighted by arrows in the right bottom panel of Fig. \ref%
{fig5}. Such evolution is a direct indication of a surprisingly high
robustness of the vortex soliton tori -- at least, for $S=1$.

In conclusion, we have found that a self-repulsive cubic nonlinearity, whose
strength grows from the center to periphery in the 3D space faster than $%
r^{3}$ (which may be implemented in BEC and in field theories with diverse
physical realizations), gives rise to robust vortex-soliton tori -- at
least, those with topological charges $S=1$ and $2$. Unstable solutions with
$S\geq 2$ split into rotating states with nested single-charged vortices. We
have found that application of a moderately strong torque sets the vortex
solitons into gyroscopic precession, with the axle moving along a conical
surface. Strong torques heavily deform the initial states, but, even under
such perturbations, the vortex-soliton tori gradually restore themselves,
with the axle directed along the total angular momentum. To the best of our
knowledge, such gyroscopic dynamics of vortex solitons have not been
encountered earlier in any comparable model.

\begin{acknowledgments}
B.A.M. appreciates hospitality of ICFO. The work of R.D. and B.A.M.
was supported, in a part, by the Binational (US-Israel) Science
Foundation through grant No. 2010239, and by the German-Israel
Foundation through grant No. I-1024-2.7/2009. R.D. and T.M.
acknowledge support provided by the Deutsche Forschungsgemeinschaft
(DFG) via the Research Training Group (GRK 1464).
\end{acknowledgments}

\end{document}